\documentclass[fleqn,12pt,twoside]{article}
\usepackage[headings]{espcrc1}

\readRCS
$Id: PhysicsProcedia2009.tex,v 1.0 2009/10/5 11:22:11 spepping Exp $
\usepackage{graphicx}
\newcommand{\captionset}{\addtolength{\leftskip}{1.1cm}\addtolength{\rightskip}{1.2cm}
 \small\textsf}

\newcommand{\D}{\mathrm{d}}
\newcommand{\e}{\mathrm{e}}

\newcommand{\kb}{k_{\mathrm{B}}}
\newcommand{\kbt}{k_{\mathrm{B}}T}

\newcommand{\be}{\begin{equation}}
\newcommand{\ee}{\end{equation}}
\newcommand{\bea}{\begin{eqnarray}}
\newcommand{\eea}{\end{eqnarray}}

\newcommand{\AmS}{{\protect\the\textfont2
  A\kern-.1667em\lower.5ex\hbox{M}\kern-.125emS}}

\title{Revisiting the Poisson-Boltzmann theory: charge surfaces, multivalent ions and inter-plate forces}

\author{Dan Ben-Yaakov\address[TAU]{Raymond and Beverly Sackler School of Physics and Astronomy, Tel Aviv
University, Ramat Aviv 69978, Tel Aviv, Israel},
       David Andelman\addressmark
        }

\runtitle{Boundaries and multivalency effects on inter-plate
forces} \runauthor{D. Ben-Yaakov and D. Andelman}

\begin{document}

% typeset front matter
\maketitle

\begin{abstract}
The Poisson-Boltzmann (PB) theory is extensively used to gain insight on
charged colloids and biological systems as well as to elucidate
fundamental properties of intermolecular forces. Many works were
devoted in the past to study  PB related features and to confirm
them experimentally. In this work we explore the properties of
inter-plate forces in terms of different boundary conditions.
We treat the cases of constant surface charge, constant surface
potential and mixed boundaries. The interplay between
electrostatic interactions,  attractive counter-ions release, and
 repulsive van 't Hoff contribution are discussed separately for each case.
Finally, we discuss how the crossover between attractive and
repulsive interactions for constant surface charge case is
influenced by the presence of multivalent counter-ions, where it is
shown that the range of the attractive interaction grows with the
valency.
\end{abstract}

%%%%%%%%%%%%%%%%%%%%%%%%%%%%%%%%%%%%
\section{Introduction} \label{intro}
%%%%%%%%%%%%%%%%%%%%%%%%%%%%%%%%%%%%
The first applications of Poisson-Boltzmann (PB) theory date back
to the beginning of the 20th century with the pioneering works of Gouy
and Chapman \cite{gouy,chapman}, which dealt with the counter-ions cloud
formed in the vicinity of charged plates. A decade later,
Debye and H\"{u}ckel \cite{debye} worked out a theory for
electrolyte solutions, based on the linear PB theory. Since then the
PB theory was extensively used for many basic and applied
purposes, such as the seminal DLVO  theory \cite{verwey,israelach} that explains
the stability of colloidal suspensions.

Nowadays, the PB theory is a benchmark tool for a variety of
experimental and industrial applications \cite{israelach,evans}, in
which it is being used to interpret experimental data quantitatively. It is a
vital ingredient in any attempt to explain intermolecular forces and
interactions between charged macromolecules, particles and surfaces.

However, the theory has some drawbacks, which were studied
during the last decades. Some of the refinements are related to the
mean-field limitations of the theory that does not account correctly for
correlations and fluctuations \cite{moreira2002,netz2000}. Other
modifications take into account additional interactions, such as
steric effects \cite{iglic1996,borukhov1997} and hydration
interaction \cite{burak2000}. All these refinements are motivated by
a number of cases where the
PB theory fails to explain experimental observations. For example, the
theory misses some important features when treating multivalent
ions and highly charged objects.

Nevertheless, as the collection of problems that one can treat with the PB
theory is very broad in its scope and range, we aim in the present work to further
explore several PB related cases. We focus on two main aspects. In
section~\ref{bc}, the force between two charged plates is
discussed as function of boundary type. In section~\ref{multivalent}, we extend the
discussion to multivalent ions and discuss their effect on the
crossover from attractive to repulsive pressure.

%%%%%%%%%%%%%%%%%%%%
\section{PB Equation} \label{bc}
%%%%%%%%%%%%%%%%%%%%

Consider a two-plate system occupying the space between two planar surfaces located at
$y{=}0$
and at $y{=}D$ ($y$ is the coordinate perpendicular to the plates).
The system is coupled to a bulk reservoir of salt
concentration $n_b$, assumed here to be monovalent, $z{=}1$.
The PB equation, describing the electrostatic
potential and the ionic distributions in between the plates, reads:
\be \frac{\D^2 \psi}{\D y^2}{=}\frac{8 \pi e n_b}{\varepsilon}\sinh
\left(\frac{e \psi}{\kbt}\right)\, , \ee
where $e$ is the electron charge, $\varepsilon$  the solvent
dielectric constant, $\kb$ the Boltzmann's constant, and $T$ is the
temperature. By defining a dimensionless potential $\phi\equiv
e\psi/\kbt$, the PB equation is written in a dimensionless form:
\be \frac{\D^2 \phi}{\D x^2}{=}\sinh \phi\, , \label{dPB}\ee
where $x{=}\kappa y$ is the dimensionless coordinate, rescaled by
the inverse Debye screening length $\kappa^{-1}$, defined by
$\kappa^2{=}8\pi e^2 n_b/ (\varepsilon \kbt)$.

Next, the boundary conditions on the plates can be defined in a
general manner. The boundary condition at $y{=}0$ and $y{=}d$ are given by:
\be f_0 \left(\phi(0),\phi'(0)\right){=}0\quad{\rm and}\quad
f_d \left(\phi(d),\phi'(d)\right){=}0\, ,\ee
where $d\equiv\kappa D$ is the dimensionless inter-plate spacing.
The functions $f_0$ and $f_d$ are chosen
with respect to the surfaces properties. This choice determines the
nature of the interaction between the plates as is elaborated below.

Using the solutions of the PB equation one can calculate global
thermodynamic quantities that are accessible in
experiments. In particular, it can be shown \cite{anddelman_safran}
that integrating once the PB equation gives the osmotic pressure,
$\Pi$, in terms of the electrostatic potential and its electric field, $\phi\,'$:
\be \Pi{=}{-}\frac{1}{2}\left(\frac{\D \phi}{\D x}\right)^2+\cosh\phi-1\, .\ee
Here the osmotic pressure is rescaled by $2\kbt n_b$, yielding the
dimensionless osmotic pressure $\Pi$. Note that the pressure is homogeneous
and has no dependence on the coordinate $x$. It is straightforward
to calculate the free energy per unit area $F$ (in a dimensionless units):
\be F{=}-\int_\infty^d \D x \Pi(x)\, . \label{free_energy}\ee

Although not always accurate, it is  instructive to
solve the linear PB equation
\be \frac{\D^2 \phi}{\D x^2}{=}\phi\, ,\ee
whose solution is given by $\phi{=}A\cosh x+B\sinh x\,$. The
constants $A$ and $B$ are obtained by satisfying the boundary
conditions
\bea f_0\left(A,B\right){=}0\, ,\hspace{6.1cm}\label{bc_0}\\
f_d\left(A\cosh{d} +B\sinh{d} ,A\sinh{d}+B\cosh{d}\right){=}0\label{bc_d}\, ,
\eea
and are functions of the separation $d$. The osmotic pressure depends on these two constants:
\be \Pi{=} \frac{1}{2}A^2-\frac{1}{2}B^2\, . \label{pressure}\ee

In the following we choose three different types of boundaries, and
discuss how these choices affect the nature of the interaction
between the plates.

%%%%%%%%%%%%%%%%%%%%%%%%%%%%%%%%%%%%%%%%
\subsection{Two constant charge plates}
%%%%%%%%%%%%%%%%%%%%%%%%%%%%%%%%%%%%%%%%
In many physical setups, the surface charge density can be
regarded to a very high accuracy as a constant. Therefore, many
theoretical works addressed this case in the past solving it numerically
\cite{preive1976,chan1976,preive1978,healy
1980,chan1995,stankovich1996,biseshevuel2004} or analytically
\cite{parsegian1972,behrems1999,meier2004,safran2005,benyaakov2007}.
The functions $f_0$ and $f_d$ are given by:
\bea f_0 \left(\phi(0),\phi'(0)\right){=}\phi'(0)+{\sigma_0}\hspace{0.2cm}\label{bc_0_sigma}
\\ f_d \left(\phi(d),\phi'(d)\right){=}\phi'(d)-{\sigma_d}\, ,\label{bc_d_sigma}\eea
where ${\sigma}_0$ and ${\sigma}_d$ are the two rescaled (dimensionless) surface
charge densities ${\sigma}_{0}\rightarrow 4\pi e
\sigma_{0}/({\kappa\varepsilon \kbt})\,$ and
${\sigma}_{d}\rightarrow 4\pi e \sigma_{d}/({\kappa\varepsilon
\kbt})\,$, respectively. Substituting them in eqs.~\ref{bc_0} and \ref{bc_d}, we
obtain \\ $A{{=}}({ {\sigma}_d+ {\sigma}_0\cosh d})/{\sinh d}$ and
$B{{=}}-{\sigma_0}\,$. The pressure and the free energy, given by
eq.~\ref{pressure} and \ref{free_energy}, read

\bea
\Pi{=}\frac{2 {\sigma}_0 {\sigma}_d\cosh d+ {\sigma}_0^2+ {\sigma}_d^2}{2\sinh^2 d}\,
,\\
F{=}\frac{2 {\sigma}_0 {\sigma}_d+( {\sigma}_0^2+ {\sigma}_d^2)\e^{-d}}{2\sinh d}\,
.\hspace{0.3cm} \eea
For large separations ($d\gg1$), the pressure decays exponentially
 $\Pi\simeq {\sigma}_0 {\sigma}_d\e^{-d}\,$,
where the sign of the pressure is determined by the relative sign
between $ {\sigma}_0$ and $ {\sigma}_d$ \cite{parsegian1972}. For unlike-charges
(${\sigma}_0\cdot {\sigma}_d<0$) the pressure is attractive ($\Pi>0$),
while like-charges ($ {\sigma}_0\cdot {\sigma}_d>0$) repel each
other. For small separations there is no dependence on the sign of
${\sigma}_0$ and $ {\sigma}_d$ and  $\Pi$ becomes purely
repulsive: $\Pi{=}{( {\sigma}_0+ {\sigma}_d)^2}/({2\sinh^2 d})>0\,$,
because of entropic reasons (see below).

Thus, there is a crossover from attractive to repulsive pressure for
unlike-charges. There is one exception for this finding: when
$ {\sigma}_0{=}- {\sigma}_d$ the pressure is purely attractive.
For like-charges, the pressure is monotonically positive and the
charges repel each other for all the separations.

The physical interpretation for this behavior is based on the
interplay between two mechanisms as was previously discussed ({\it e.g.},
see Ref.~\cite{benyaakov2007}). When the plates are brought closer, the
neutralizing ionic clouds increasingly overlap. Since the ionic
clouds have opposite signs, pairs of negative and positive
counter-ions gain additional entropy by staying in the bulk instead
of neutralizing the system, while the plates, in turn, neutralize
each other. The closer the plates are, the larger is the ionic cloud
overlap, and
more counter-ions are released to the bulk resulting in an
attractive pressure. However, when the magnitude of the surface
charges is dissimilar ($| {\sigma}_0|\neq| {\sigma}_d|$), a finite
amount of counter-ions are forced to stay in between the plates to
neutralize the excess of charge. The concentration of these neutralizing
counter-ions grow when the separation is reduced, leading to a
repulsive pressure (due to the van 't Hoff pressure of the
counter-ions). For small separations ($d\rightarrow 0$), this
repulsion diverges as $\Pi\sim1/d$.

For the pure antisymmetric case, where ${\sigma}_0{=}- {\sigma}_d\,$,
only the counter-ion release mechanism is observed since no
counter-ions are needed to neutralize the  plates in small
separations (there is no charge excess and the two plates completely neutralize
each other). This is the origin of the pure attractive
pressure in this case. On the other hand when $ {\sigma}_0\cdot
{\sigma}_d>0$, only the repulsive pressure, caused by
counter-ion neutralization, takes place since the clouds have similar
signs, and there is no counter-ions release. We hence conclude that
the crossover from attractive to repulsive pressure occurs only for
unlike-charges ($ {\sigma}_0\cdot {\sigma}_d<0$) with asymmetric
magnitudes, $|{\sigma}_0|\neq| {\sigma}_d|$.

In Figure~1a three curves of pressure vs. distance with constant
surface charge boundaries are shown. In addition to the simple analytical results
obtained for the linearized PB case, we solve numerically
the non-linear PB equation. Both are presented in Figure~1;
the linear case as dashed lines while the non-linear case as solid lines.
The comparison between the solid and dashed curves shows
that the solution of the linear equation agrees qualitatively with
the exact solution of the non-linear equation. The numerical
deviations between the two are small for the chosen values of $\sigma_0$
and $\sigma_d$. However, for larger surface charges the
deviations grow, and the linear solution is not sufficiently
accurate anymore.
{\linespread{1.0}
\begin{figure}[tb]\centering
 \includegraphics[width=0.43\textwidth]{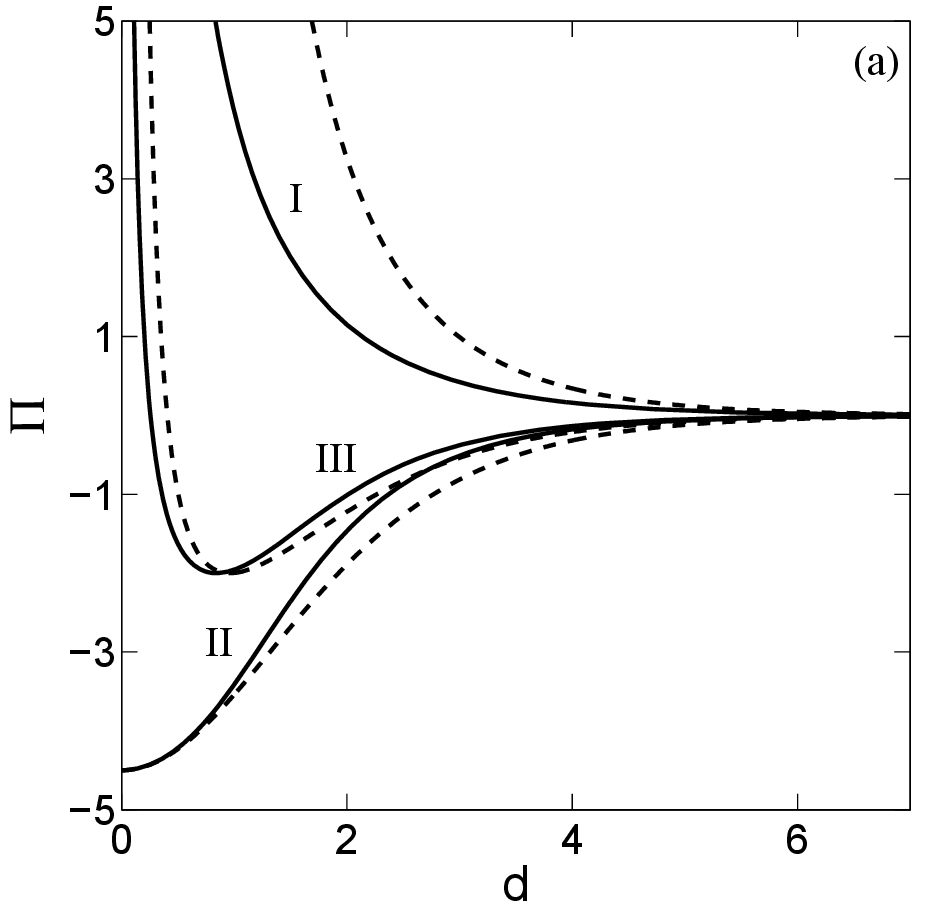}
  \includegraphics[width=0.43\textwidth]{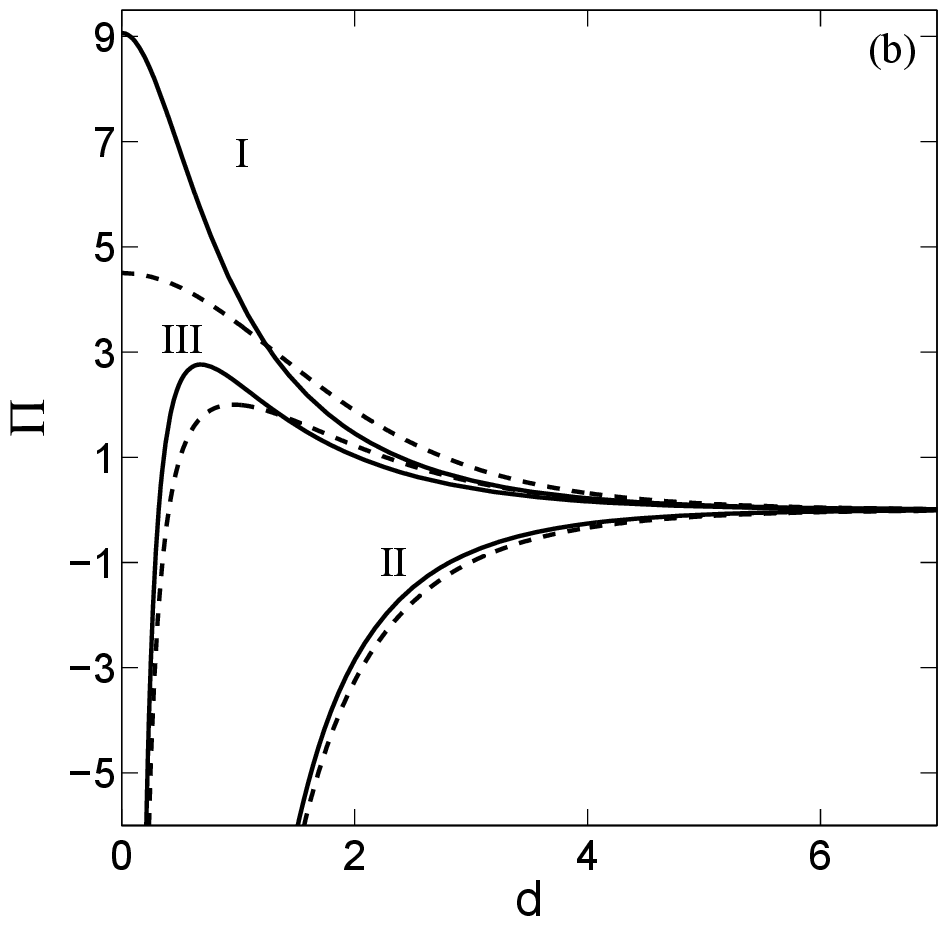}
 \caption{\captionset{Osmotic pressure $\Pi$ as function of inter-plate
 separation, $d$. The solution to the linear (dashed) and
non-linear (solid) PB equation are presented for  constant
surface charge case in (a) and  constant surface potential
in (b). Three profiles are shown for each case:  {\rm I, II}
and {\rm III}  indicate the repulsive, attractive and a crossover
profile, respectively. (a) The boundary conditions for profiles {\rm
I, II} and {\rm III} are: $\sigma_0{=}\sigma_d{=}3$,
$\sigma_0{=}{-}\sigma_d{=}3$ and $\sigma_0{=}{-}3\sigma_d/2{=}3$,
respectively. (b) The boundary conditions for  {\rm I, II} and
{\rm III} are: $\phi_0{=}\phi_d{=}3$, $\phi_0{=}{-}\phi_d{=}3$ and
$\phi_0{=}3\phi_d/2{=}3$, respectively.}} \label{fig1}
\end{figure}
}
%

%%%%%%%%%%%%%%%%%%%%%%%%%%%%%%%%%%%%%%%%%%%%
\subsection{Two constant surface potentials}
%%%%%%%%%%%%%%%%%%%%%%%%%%%%%%%%%%%%%%%%%%%%
The functions $f_0$ and $f_d$ are given here
by:
\bea f_0 \left(\phi(0),\phi'(0)\right){=}\phi(0)-{\phi_0}
\hspace{0.2cm}\\ f_d \left(\phi(d),\phi'(d)\right){=}\phi(d)-{\phi_d}\, ,\eea
where $\phi_0$ and $\phi_d$ are dimensionless potentials. Obtaining
the constants $A$ and $B$ from these two relations, $A{=}\phi_0\,$
and $B{=}({\phi_d-\phi_0\cosh d})/{\sinh d}\,$, leads to the dimensionless
osmotic
pressure $\Pi$ and free energy $F$:

\bea
\Pi{=}\frac{2\phi_0\phi_d\cosh d-\phi_0^2-\phi_d^2}{2\sinh^2 d}\,
,\\
F{=}\frac{2\phi_0\phi_d-(\phi_0^2+\phi_d^2)\e^{-d}}{2\sinh d}\,
.\hspace{0.3cm} \eea
These expressions are almost identical to the constant surface
charge case (except for the sign), and indeed, for large
separations the relative sign of $\phi_0$ and $\phi_d$ determines
the pressure: $\Pi\simeq\phi_0\phi_d\e^{-d}\,$, similarly to the
large separation behavior of the constant charge case. However, for
small separations, the pressure is different:
$\Pi{\simeq}-{(\phi_0-\phi_d)^2}/2{\sinh^2 d}\,$, yielding a pure
attractive pressure, except for $\Pi{=}0$ when $\phi_0{=}\phi_d$.
Unlike the constant charge case, in the constant potential case the
counter-ion concentration remains constant near each plate, but the
effective surface charge diverges when the plates are brought
closer together. This results in a diverging electrostatic attraction that
governs the pressure for small $d$.

Another remark about the constant potential case is that the crossover from
repulsive to attractive pressure prevails only for like-potentials
($\phi_0\cdot \phi_d{>}0$). For unlike-potentials ($\phi_0\cdot
\phi_d{<}0$) the pressure is purely attractive, while for the
special symmetric case, $\phi_0{=}\phi_d\,$, the pressure is
purely repulsive.

In Figure~1b we show three curves of pressure vs. distance for three
different surface potential values (attractive, repulsive and crossover). Note
that the attractive profile diverges, while the repulsive one
saturates and is different than the constant surface charge case, where
the repulsive one diverges and the attractive one saturates. This
is due to the different constraints that the boundaries apply on the
ionic solution. Also note that the dominant deviations of the exact
non-linear PB solution from the linearized one are large for small
inter-plate separations, due to the large overlap between the two induced counter-ion clouds.

%%%%%%%%%%%%%%%%%%%%%%%%%%%
\subsection{The mixed case}
%%%%%%%%%%%%%%%%%%%%%%%%%%%

The third case  is a mixed boundary condition, where the plate at
$y{=}0$ has a constant potential while the one at $y{=}D$ has a
constant charge. The functions $f_0$ and $f_d$ are given by: $f_0
\left(\phi(0),\phi'(0)\right){=}\phi(0)-{\phi_0}\,$, and $f_d
\left(\phi(d),\phi'(d)\right){=}\phi'(d)-{ {\sigma}_d}\,$.
Substituting the solution for $\phi$ in these two relations, the constants $A$ and $B$
are obtained: $A{=}\phi_0$ and $B{=}({ {\sigma}_d-\phi_0\sinh
d})/{\cosh d}\,$. The pressure $\Pi$ and the free energy $F$ are
given by:
\bea
\Pi{=}\frac{2\phi_0 {\sigma}_d\sinh d+\phi_0^2- {\sigma}_d^2}{2\cosh^2 d}\,
,\\
F{=}\frac{2\phi_0 {\sigma}_d+(\phi_0^2- {\sigma}_d^2)\e^{-d}}{2\cosh d}\,
.\hspace{0.2cm} \eea
The large separation the pressure behaves as $\Pi{\simeq}\phi_0
{\sigma}_d\e^{-d}\,$, similarly to the previous two cases where the
nature of the pressure is determined by the relative sign
($\phi_0\cdot  {\sigma}_d$). The similarity between the three cases
for large $d$ is due to the decoupling between the plates,
effectively leading to two isolated planes with no
mutual effect on their surface potential or charge. On the other hand,
examining the
small separation limit, we obtain\\ $\Pi{\simeq}{(\phi_0^2-
{\sigma}_d^2)}/(2{\cosh^2 d})\,$ where the sign of the pressure
depends on the magnitudes of $\phi_0$ and $ {\sigma}_d$, unlike the
former cases. When $| {\sigma}_d|{>}|\phi_0|$ the pressure is
attractive, while in the opposite case, $| {\sigma}_d|{<}|\phi_0|$,
the plates repel each other. The competition between the
counter-ions release, the charge neutralization, and constant
ion density at the constant potential surface leads to this
versatile behavior. Moreover, in contrast to the former cases,
there is no divergence here at small separations, and the pressure
saturates as $d\rightarrow 0$. Note the special case where $|
{\sigma}_d|{=}|\phi_0|$ that leads to vanishing of the pressure
 as $d\rightarrow 0$.

In this section we demonstrate how versatile can be the electrostatic
pressure between two plates in the framework of the linear PB theory. We
 note that the three cases treated here have the simplest
boundary conditions one can consider. In physical realizations, one should
often take also into account  the charge regulation mechanisms \cite{ninham1971},
where the surface properties are determined by additional surface
parameters such as the ionic dissociation degree. In these cases the functions
$f_0$ and $f_d$  have a more complex form, that describes the
relation between $\phi$ and $\phi'$ at $x{=}0$ or $x{=}d$ due to surface
activity. Although our results are derived for the linearized PB case,
we expect similar conclusions to hold for the general non-linear PB case as well.

%%%%%%%%%%%%%%%%%%%%%%%%%%%%%%%%%%%%%%%%%%%%%%%%%%%%%%%%%%%%%%%%
\section{Attraction to Repulsion Crossover with Multivalent Salts}\label{multivalent}
%%%%%%%%%%%%%%%%%%%%%%%%%%%%%%%%%%%%%%%%%%%%%%%%%%%%%%%%%%%%%%%%
We consider a two-plate system with unlike-charges,
${\sigma_0}{<}0$ and ${\sigma}_d{>}0$ in the presence of ionic
solution with multivalent salt. First we treat the case of a symmetric
salt ($z{:}z$) and then consider mixtures of two salts: a symmetric
monovalent salt (1{:}1) combined with an asymmetric multivalent salt ($1{:}z$).

%%%%%%%%%%%%%%%%%%%%%%%%%%%%%%%%%%%%%%%%%%%%%%
\subsection{Multivalent symmetric salt ($z{:}z$)}
%%%%%%%%%%%%%%%%%%%%%%%%%%%%%%%%%%%%%%%%%%%%%%
In this case the dimensionless PB equation of eq.~\ref{dPB} is generalized and reads:
\begin{equation}
\frac{\D^2 \phi}{\D x^2}{=}z\sinh{(z\phi)}\,.
\end{equation}
where $z$ is the ionic valency. The boundary conditions at
$x{=}0$ and $d$ are given by \be f_0
\left(\phi(0),\phi'(0)\right){=}\phi'(0)+{\sigma_0}/z\quad{\rm and
}\quad f_d \left(\phi(d),\phi'(d)\right){=}\phi'(d)-{\sigma_d}/z\,.
\ee In this form the equation is almost identical to the monovalent
salt case (eq.~\ref{dPB}). Therefore, the attraction to repulsion condition
of the monovalent case (as was discussed in Ref.~\cite{benyaakov2007}) can be used:
\be
\Pi {=} \left\{
\begin{array}{l l}
  <0 & \e^{-\sqrt{z}d}<{\gamma_0}/{\gamma_d}<\e^{\sqrt{z}d}\quad {\rm and}\quad {\sigma}_0\cdot{\sigma}_d<0\\
   >0 &  {\rm otherwise}\\
\end{array} \right.
\ee

where $\gamma_{0,d}{=}\sqrt{\left(2\sqrt{z}/{\sigma}_{0,d}\right)^2+1}
-\left|2\sqrt{z}/{\sigma}_{0,d}\right|$.

%%%%%%%%%%%%%%%%%%%%%%%%%%%%%%%%%%%%%%%%%%%%%%%%%%%%%%%%%%%%%%%%%%%%%%%%%%%%%%%%%%%
\subsection{Mixtures of electrolytes}
%%%%%%%%%%%%%%%%%%%%%%%%%%%%%%%%%%%%%%%%%%%%%%%%%%%%%%%%%%%%%%%%%%%%%%%%%%%%%%%%%%%

Next we consider a binary salt mixture: the first
is an asymmetric salt ($1{:}z$) for which the bulk concentration of the
multivalent cations and monovalent anions are $n_{b}^{z}$ and $z
n_{b}^{z}\,$, respectively. The second is a monovalent salt (1{:}1) for
which the bulk concentration $n_{b}\,$. The dimensionless PB equation now reads:
\begin{equation}
\frac{\D^2 \phi}{\D
x^2}{=}\sinh{\phi}-\frac{z\alpha}{2}\left(e^{-z\phi}-e^{\phi}\right)\,.
\end{equation}
where $\alpha{=}n_{b}^{z}/n_{b}$ is the ratio between the bulk
densities of the multivalent and monovalent cations.
This equation depends explicitly on the parameters $\alpha$ and $z$,
while all the other physical parameters are taken into account by
the dimensionless boundary conditions (eqs.~\ref{bc_0_sigma} and
\ref{bc_d_sigma}). The first integration of this equation gives the
dimensionless pressure in between the plates:
\begin{equation}
\Pi_{\rm in}{=}-\frac{1}{2}\phi'\,^2+\cosh\phi+\frac{\alpha}{2}
e^{-z\phi}+\frac{z\alpha}{2} e^{\phi}
\end{equation}
The pressure in the outer region is calculated by taking $\phi{=}\phi'{=}0$:
\begin{equation}
\Pi_{\rm out}{=}1+\frac{\alpha}{2}+\frac{z\alpha}{2}
\end{equation}
In order to calculate the attraction/repulaion crossover line we examine the point in the parameter
space where the net pressure on the plates $\Pi{=}\Pi_{\rm
in}-\Pi_{\rm out}$ vanishes, yielding the following
relation between $\phi$ and $\phi'$:
\begin{equation}
\phi'{=}\sqrt{2(\cosh\phi-1)+\alpha(e^{-z\phi}-1)+z \alpha (e^{\phi}-1)}
\end{equation}
By solving this first-order differential equation, the
boundary line between attraction and repulsion regions in the
($d$\,,\,$|{\sigma}_d/{\sigma}_0|$) plane can be obtained.

In the limit of low salt densities, one can assume that
the concentration of the multivalent counterions
is much larger than that of the monovalent ones
due to entropy considerations. Hence, we
can solve the counter-ions only equation,
{\linespread{1.0}
\begin{figure}[tb]\centering
 \includegraphics[width=0.46\textwidth]{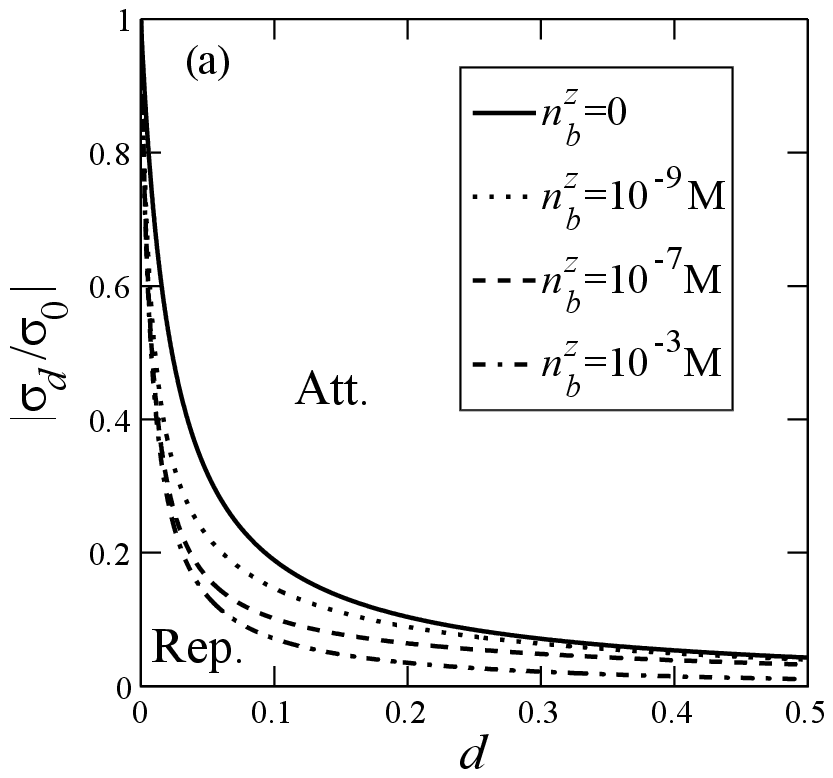}
  \includegraphics[width=0.435\textwidth]{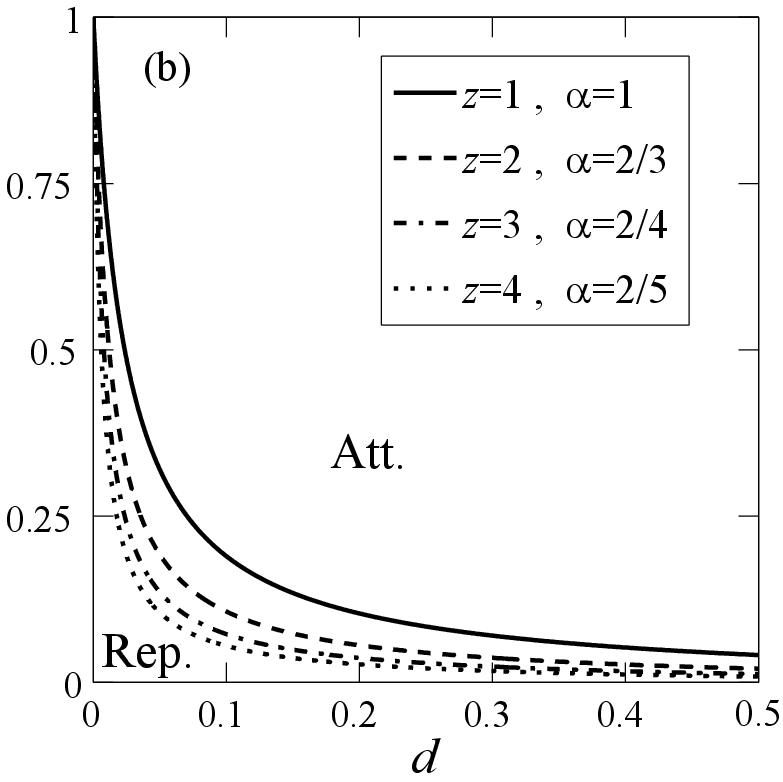}
 \caption{\captionset{Crossover between attraction and repulsion for binary salt mixtures. (a) Different values  of the
 multivalent cation concentration $n_b^z$ (see box). The
monovalent salt concentration is $n_{b}{=}10^{-3}$\,M and the negative
surface charge density is $\sigma_0{=}-e/100$\AA$^{-2}$.  The
valency of the cations is $z{=}+3$. (b) Varying the multivalent
cation valency, $z$. The negative surface charge density $\sigma_0$
is the same as in (a). The overall ion bulk concentration,
$n_{t}{=}2 n_b+(1+z)\alpha n_b{=}4n_b\,$, is kept constant for all
the curves, yielding a constant reference pressure in the outer region.}} \label{fig2}
\end{figure}
}
%
%such that the reference pressure in the outer region is
%kept constant. Thus, the effect, being examined, is how the valency
%changes the crossover.
and derive a similar condition for low salt limit as was derived in Ref.~\cite{benyaakov2007}:
\begin{equation}
\label{no_salt_cond}
\left|\frac{1}{\sigma_{0}}-\frac{1}{|\sigma_{d}|}\right|<\frac{1}{{\sigma}^*}\,
,
\end{equation}
where ${\sigma}^*{=}\varepsilon\kbt/(2\pi e D)$ is a
characteristic charge surface density that depends on $D$. This
condition has no dependence on the salt concentration, so that the
multivalent and monovalent cases have the same behavior in this
limit.

In Figure~2a we show the effect of adding multivalent salt to the
solution. It is evident from the plot that the attraction region
grows with adding multivalent salt. In Figure~2b the effect of
increasing the valency of the multivalent cations is examined. In the plot
it can be seen  that the attraction region is increased for
higher valencies. It is interesting to note that the crossover
distance $d$ scales inversely with the valency for constant asymmetry ratio
$|\sigma_d/\sigma_0|$.

%%%%%%%%%%%%%%%%%
\section{Summary}
%%%%%%%%%%%%%%%%%
In this work we explored a variety of PB theory results.
In particular, we focused on several types of charge boundaries that affect
the nature of the pressure between two plates, as well as how multivalent ions
influence the crossover between attractive to repulsive inter-plate pressure.

It is evident from this work that solutions of the linear PB
equation can serve as a qualitative tool for understanding the
interaction. However, when interpreting experimental data, one
should solve the non-linear PB theory in order to have better
numerical accuracy, especially for small inter-plate separation, $d$.

Finally, we would like to stress that the surfaces cannot always
be regarded as having either a constant surface charge or potential. In quite a number of
experimental setups the surface properties depend explicitly on the inter-plate separation.
For example, the degree of dissociation of surface charge groups can vary with $d$.
In order to take this effect into account one
should use boundary conditions that are determined  self-consistently as function of separation
\cite{ninham1971}.

%%%%%%%%%%%%%%%%%%%%%%%%
\section{Acknowledgment}
%%%%%%%%%%%%%%%%%%%%%%%%

This work was initiated during a visit to the physics department
of Kyoto University. We thank K. Yoshikawa for his hospitality and numerous discussions.
Support from the Israel
Science Foundation (ISF) under Grant No. 231/08 and the U.S.-Israel
Binational Foundation (BSF) under Grant No. 2006/055 is gratefully acknowledged.

%%%%%%%%%%%%%%%%%%%%%%%%%%

%%%%%%%%%%%%%%%%%%%%%

\end{document}